\documentclass[prb,showpacs, notitlepage]{revtex4}

\usepackage{graphicx}
\usepackage{amsfonts}
\usepackage{color}

\usepackage{epstopdf}
\usepackage{latexsym}
\usepackage{amssymb}
\usepackage{amssymb}

\usepackage[center]{subfigure}

\begin{document}

\title{Can gravity be repulsive?}

\author{M.-N. C\'{e}l\'{e}rier$^{1}$}
\email{marie-noelle.celerier@obspm.fr}

\author{N. O. Santos$^{1,2,3}$}
\email{n.o.santos@qmul.ac.uk}
 
\author{V. H. Satheeshkumar$^{3,4}$}
\email{vhsatheeshkumar@gmail.com}

\affiliation{
$^{1}$LERMA, CNRS UMR8112, Universit\'e Paris-Sorbonne, Universit\'e Pierre et Marie Curie (UPMC Paris 6), Universit\'e de Cergy-Pontoise, Observatoire de Paris-Meudon 5, Place Jules Janssen, F-92195 Meudon Cedex, France.\\
$^{2}$School of Mathematical Sciences, Queen Mary,
University of London, London E1 4NS, UK.\\
$^{3}$Departamento de F\'{\i}sica Te\'orica, Instituto de F\'{\i}sica,
Universidade do Estado do Rio de Janeiro,
Rio de Janeiro, RJ 20550-900, Brazil.\\
$^{4}$Departamento de Astronomia, Observat\'{o}rio Nacional, Rio de Janeiro, RJ 20921-400, Brazil.
}

\date{\today}
   
 \newcommand{\bq}{\begin{equation}}
 \newcommand{\eq}{\end{equation}}
 \newcommand{\bqn}{\begin{eqnarray}}
 \newcommand{\eqn}{\end{eqnarray}}
 \newcommand{\nb}{\nonumber}
 \newcommand{\lb}{\label}


\begin{abstract}
General Relativity has had tremendous successes on both theoretical and experimental fronts for over a century by now. However, the theory contents are far from being exhausted. Only very recently, with gravitational wave detection from colliding black holes, have we started probing gravity behavior in the strongly non-linear regime. Even today, black hole studies keep revealing more and more paradoxes and bizarre results. In this paper, inspired by David Hilbert's startling observation, we show that, contrary to the conventional wisdom, a freely falling test particle feels gravitational repulsion by a black hole as seen by an asymptotic observer. We dig deeper into this relativistic gravity surprising behavior and offer some explanations. 
\end{abstract}


\maketitle

\section{Introduction} 

More than a century ago, on November 25, 1915, Albert Einstein \cite{Einstein:1915ca} wrote down General Relativity (GR) field equations in their final form and soon afterwards, on January 13, 1916, Karl Schwarzschild \cite{Schwarzschild} discovered Einstein field equation first solution. Using these results, while studying particle radial motion in a spherically symmetric static gravitational field, David Hilbert \cite{Hilbert} noticed on December 23, 1916, that the acceleration of a massive particle freely falling in a gravitational field,  as seen by an asymptotic observer, changes sign when the velocity reaches a given value. It is as though the gravitational field becomes repulsive. 
Since this result of Hilbert cannot be found in any textbook on General Relativity, we are clearly explaining it in this pedagogical essay.  

It is fair to say that most of the motion we see in the Universe is a free fall in a gravitational field. There are three vantage points for looking at a particle falling into a black hole. 
 The first one is an inertial asymptotic observer whom we call \textit{far-away observer} (Wheeler's \textit{Schwarzschild bookkeeper}~\cite{Taylor-Wheeler}) who, by definition, is located infinitely far away from the black hole and is sufficiently massless not to exert any gravitational field that could affect the measurements.
 The second viewpoint is that of the freely falling particle itself which we call \textit{freely-falling observer} (Wheeler's \textit{free-float frame}~\cite{Taylor-Wheeler}). 
 The third observer is the one at a finite distance from the black hole who is referred to as \textit{finite-distance observer}(Wheeler's \textit{shell observer}~\cite{Taylor-Wheeler}).

 As seen by the far-away observer, the particle falling into the black hole never crosses the horizon but asymptotically approaches it. This is a standard textbook statement but it hides a subtle detail. We know that when a particle starts falling freely towards a black hole, the velocity increases as it approaches the gravitational source. But we also know that, from the viewpoint of an asymptotic observer, the freely falling particle never makes it to the event horizon. This means that the particle should slow down at some finite distance from the black hole and eventually cease to move at the horizon, albeit this takes an infinite time in the far-away observer's clock. In other words, the freely falling particle decelerates (the acceleration direction is positive (radially away from the gravitating source) while the velocity is negative) i.e., it is repelled by gravity as seen by the far-away observer. One must be careful about the terminology. Repulsion here does not mean that the freely falling particle is bounced off the horizon. For a historical perspective on this topic, we refer the reader to a review by Spallicci~\cite{Spallicci}.  

\section{Falling into a Schwarzschild black hole}

The Schwarzschild metric in a far-way observer's standard coordinate set up  is given by
\bqn
ds^2 &=& \left( 1-\frac{2\mu}{r} \right) c^2 dt^2 - \left( 1-\frac{2\mu}{r} \right)^{-1} dr^2 
 - r^2 (d\theta^2 + sin\theta^2 d\phi^2)
\eqn
with $$ \mu = \frac{G M}{c^2},$$ 
where $M$ is the black hole mass, and the $G$ and $c$ constants have their usual meanings. 

For simplicity, we consider a massive test particle radial motion in a Schwarzschild black hole equatorial plane. Then this particle timelike geodesic  motion obeys the following,
\bqn
\label{2}
\frac{dt}{d\tau}  &=& \left( 1-\frac{2\mu}{r} \right)^{-1}, \\
\label{3}
\frac{dr}{d\tau}  &=& \mp \left( \frac{2\mu}{r} \right)^{\frac{1}{2}}.
\eqn
Here the minus-or-plus sign in the second equation stands for infalling or outgoing test particles respectively. So now we can find the  freely falling particle velocity as measured by every observer. 
\begin{itemize}

\item The far-away observer would measure this velocity to be
\bqn
\frac{d r}{dt} = - \left( \frac{2 \mu}{r} \right)^{\frac{1}{2}} \left( 1-\frac{2\mu}{r} \right) ,
\eqn
which is zero on the horizon. 

\item The finite-distance observer measures the proper time interval as $ d t' = \left( 1-\frac{2\mu}{r} \right)^{\frac{1}{2}} \,dt,$ and the proper distance as $ d r' = \left( 1-\frac{2\mu}{r} \right)^{-\frac{1}{2}} \,dr.$ Hence the freely falling particle velocity as recorded by this observer is 
\bqn
\frac{d r'}{dt'} = - \left( \frac{2 \mu}{r} \right)^{\frac{1}{2}}.
\eqn
After restoring the constants, its absolute value is equal to $c$ on the horizon. This means the shell observer sees the test particle velocity approaching light velocity as it nears the horizon, which is quite opposite to a far-away observer's measurement.

\item Of course, the freely-falling observer is at rest in the float-frame throughout the course of motion, even when passing through the horizon, as long as the tidal forces do not kick in~\cite{Moore:2013sra}. 
\end{itemize}

The test particle radial geodesic motion equations  in a Schwarzschild field are given by
\bqn
\label{6}
\frac{d^2 r}{d\tau^2} &=& - \left( \frac{\mu}{r^2} \right) \left( 1 - \frac{2 \mu}{r} \right) \left( \frac{d t}{d \tau} \right)^2 
+ \left( \frac{\mu}{r^2} \right) \left( 1 - \frac{2 \mu}{r} \right)^{-1} \left( \frac{d r}{d \tau} \right)^2, \\
\label{7}
\frac{d^2 t}{d\tau^2} &=& -2 \left( \frac{\mu}{r^2} \right) \left( 1 - \frac{2 \mu}{r} \right)^{-1} \frac{d t}{d \tau} \frac{d r}{d \tau}.
\eqn
Upon substituting Eq.(\ref{2}) and Eq.(\ref{3}) in Eq. (\ref{6}), we obtain 
\bqn
\frac{d^2 r}{d\tau^2} &=& - \frac{\mu}{r^2}.
\eqn
For a freely falling test particle whose proper time is $\tau$, the following relation holds,
\bqn
\frac{dr}{d\tau} = \left( \frac{dr}{dt} \right) \frac{dt}{d\tau}.
\eqn
Further, we have 
\bqn
\frac{d^2 r}{d\tau^2} = \frac{d^2 r}{dt^2} \left( \frac{dt}{d\tau} \right)^2 +  \frac{dr}{dt} \, \frac{d^2 t}{d\tau^2},
\eqn
where $\frac{dr}{dt}$ and $\frac{d^2r}{dt^2}$ are the freely falling particle velocity and acceleration as recorded by the far-way observer. Substituting $\frac{d^2 r}{d\tau^2}$ and $\frac{d^2 t}{d\tau^2}$ from the geodesic equations, we get 
\bqn
\label{11}
\frac{d^2 r}{dt^2} = \left( \frac{\mu}{r^2} \right) \left[ \frac{3}{ \left( 1-\frac{2\mu}{r} \right)} \left( \frac{dr}{dt} \right)^2 - \left( 1-\frac{2\mu}{r} \right) \right]. \nb\\
\eqn
From this, we can see that the freely falling particle acceleration is positive if 
\bqn
\frac{d r}{dt} > \frac{1}{\sqrt{3}} \left( 1-\frac{2\mu}{r} \right).
\eqn
This agrees with Hilbert's original result~\cite{Hilbert} which was reproduced much later on by McGruder~\cite{McGruder}.
Eq.(\ref{11}) can be simplified to give
\bqn
\frac{d^2 r}{dt^2} = - \left( \frac{\mu}{r^2} \right) \left( 1-\frac{2\mu}{r} \right) \left( 1 - \frac{6\mu}{r}  \right), \nb\\
\eqn

 We infer the following from this result. Here, we have restored $c$ to make the situation easier to understand. 
\begin{itemize}

\item Any particle with velocity (as measured by the far-away observer) greater than $\frac{c}{\sqrt{3}}$ at any point in the Schwarzschild field is repelled by gravity. 

\item If a test particle starts from rest at infinity (where spacetime is flat for all practical purposes) and because of the black hole presence, it freely falls into the latter. During the free fall, the particle accelerates only till the velocity (as measured by the far-away observer) reaches $\frac{-2c}{3\sqrt{3}}$. At that point, its acceleration switches sign and the test particle starts decelerating and eventually (after an infinite time as measured by the far-away observer) comes to a halt at the horizon.

\item If a test particle is launched toward a black hole with an initial velocity equal to or greater than $\frac{c}{\sqrt{3}}$, it decelerates all the way to the horizon. 

\item When a light ray is shined at a black hole, it approaches it with ever decreasing velocity only to stop at the horizon. 
\end{itemize}

\begin{table*}[htp]
\label{default}
\caption{Three views of falling into a Schwarzschild black hole}
\begin{center}
\begin{tabular}{l c c c}
\hline
\hline
 Quantity &  Far-away observer & Finite-distance observer & Freely-falling observer\\ 
\hline
Existence & Far-away & Outside Horizon & Everywhere \\
Time & $dt$ & $dt' = \left( 1 - \frac{2\mu}{r}  \right)^{\frac{1}{2}} dt$ & $d\tau$ \\
Distance & $dr$ & $dr' = \left( 1 - \frac{2\mu}{r}  \right)^{-\frac{1}{2}} dr$ & $c\,d\tau$ \\
Velocity & $-\left( \frac{2\mu}{r}  \right)^{\frac{1}{2}} \left( 1 - \frac{2\mu}{r}  \right)$ & $-\left( \frac{2\mu}{r}  \right)^{\frac{1}{2}}$ & 0 \\
Maximum Velocity  & $-\frac{2c}{3\sqrt{3}} $ & $-c$ & 0 \\
Velocity at horizon & {0} & {$\rightarrow c$} & {0} \\
Acceleration  & $- \left( \frac{\mu}{r^2}  \right) \left( 1 - \frac{2\mu}{r}  \right) \left( 1 - \frac{6\mu}{r} \right) $ & $- \left( \frac{\mu}{r^2}  \right) \left( 1 - \frac{2\mu}{r}  \right)^{\frac{1}{2}} $ & 0 \\
Acceleration at horizon & 0 &  0 &  0 \\
Repulsion  & {$r = 6 \mu $} & {Never} & {Never} \\
\hline
\hline
\end{tabular}
\end{center}
\end{table*}

\section{Conclusions}

Given general relativity overwhelming successes for over a century, what is amazing is that the theory still has some surprises  for us. 
This kind of gravity repulsive behavior is commonly known in two different scenarios albeit in completely different contexts. Firstly, this is reminiscent of the cosmological term  in the weak field limit, 
where ``Newtonian'' gravity is described by, 
\bq
\nabla^2 \Phi = 4 \pi G \rho - \Lambda c^2.
\eq
Hence, the acceleration due to a spherical mass $M$ gravity is given by 
\bq
a = - |\overrightarrow{\nabla} \Phi| = -\frac{GM}{r^2} \, + \frac{c^2 \Lambda r}{3}.  
\eq
The repulsive effect comes from the second term containing the positive cosmological constant $\Lambda$. 
Secondly, one may also notice that this repulsion is akin to the centrifugal force due to angular momentum in  Newtonian mechanics, where the acceleration of a particle orbiting around a spherical mass $M$ is given by
\bq
a =  -\frac{GM}{r^2}\,\, + \frac{h^2}{r^3}.  
\eq
Here the repulsive effect comes from the second term containing the angular momentum $h$.  
But our case exhibits no cosmological constant and is a pure radial motion. This peculiarity is specific to relativistic gravity. 

Acceleration enjoys a special status in Newtonian mechanics where space and time are absolute concepts. It has a prominent role in distinguishing between inertial and non-inertial frames. Acceleration and force are invariant under Galilean transformations and hence all observers agree on them. Since acceleration is an experimentally measurable quantity, one has to handle it with care. For more on this, we refer the reader to Petkov~\cite{Petkov}. One may say that this black hole gravitational field strange behavior in GR is simply a coordinate effect. But we know from GR general coordinate (diffeomorphism) invariance that any observer is as good as any other. What one measures in one's reference frame is as physical as the effect measured in a different frame. But what is important is that all of them are governed by the same laws of physics, i.e., Einstein field equations. These bizarre result origin lies in the fact that the quantities measured by different observers we are comparing are neither Lorentz scalars nor gauge-invariant. 
Perhaps late John Wheeler would have said about this puzzling result:``surprise without surprise".

\section*{Acknowledgments}
We are grateful to Maria de F\'{a}tima Alves da Silva for facilitating this work. VHS is supported by Ci\^{e}ncia Sem Fronteiras, No. A045/2013 CAPES and PCI-DB Fellowship from CNPq at the initial stages of this work, and is currently supported by FAPERJ though Programa P\'{o}s-doutorado Nota 10. VHS would like to thank Anzhong Wang and Sofiane Faci for several enlightening discussions, and Jailson Alcaniz for hospitality at Observat\'{o}rio Nacional.



\begin{thebibliography}{nbound}

\bibitem{Einstein:1915ca} 
  A.~Einstein,
  Sitzungsber.\ Preuss.\ Akad.\ Wiss.\ Berlin (Math.\ Phys.\ ) {\bf 1915}, 844 (1915).

\bibitem{Schwarzschild} 
  Schawarzschild, K., 
  {\em Sitzber. Deut. Akad. Wiss. Berlin, Kl. Math. Phys. Tech.} 189, (1916). 
  For English version see: {\em Gen. Rel. Grav.}, 35, 951--959, (2003).
    
\bibitem{Hilbert} 
  Hilbert, D., 
  {\em Nachrichten König. Gesell. Wiss. G\"{o}ttingen, Math.-Phys. Kl.}, 53 (1917); 
  Hilbert, D., 
  {\em Math. Ann.} 92, 1 (1924).
  For English version see Hilbert, D.,  pp 1017--1038 in {\em The Genesis of General Relativity, Volume 4 -   Gravitation in the Twilight of Classical Physics: The Promise of Mathematics}, edited by J\"{u}rgen Renn and Matthias Schemmel (Springer, Dordrecht. 2007).

\bibitem{Taylor-Wheeler} 
  E.~F.~Taylor and J.~A.~Wheeler,
  {\it Exploring Black Holes} 
  Addison Wesley Longman, San Francisco, (2000).

\bibitem{Spallicci} 
  A.~Spallicci,
  Fundam.\ Theor.\ Phys.\  {\bf 162}, 561 (2011)
  [arXiv:1005.0611 [physics.hist-ph]].

\bibitem{Moore:2013sra} 
  D.~G.~Moore and V.~H.~Satheeshkumar,
  Int.\ J.\ Mod.\ Phys.\ D {\bf 22}, 1342026 (2013)
  [arXiv:1305.7221 [gr-qc]].

\bibitem{McGruder} 
  C.~H.~McGruder,
  Phys.\ Rev.\ D {\bf 25}, 3191 (1982).
  
\bibitem{Petkov} 
   V.~Petkov, 
  ``Physics as Spacetime Geometry,''
   in \textit{Springer Handbook of Spacetime}, edited by A. Ashtekar and V. Petkov (Springer, 2014) pp 141--163.
  
  
\end{thebibliography}
\end{document}